\documentclass[sigconf,screen]{acmart}

\usepackage{amsmath}     
\usepackage{algorithmic}
\usepackage{graphicx}
\usepackage{textcomp}
\usepackage{bmpsize}     
\usepackage{xcolor}
\usepackage{lipsum}
\usepackage{booktabs}
\usepackage{multirow}
\usepackage{tabularx}
\usepackage{tcolorbox}
\usepackage{bm}
\usepackage{tablefootnote}
\usepackage{float}
\usepackage{balance}
\copyrightyear{2026}
\acmYear{2026}
\setcopyright{cc}
\setcctype{by}
\acmConference[FSE Companion '26]{34th ACM Joint European Software Engineering Conference and Symposium on the Foundations of Software Engineering}{July 05--09, 2026}{Montreal, QC, Canada}
\acmBooktitle{34th ACM Joint European Software Engineering Conference and Symposium on the Foundations of Software Engineering (FSE Companion '26), July 05--09, 2026, Montreal, QC, Canada}
\acmDOI{10.1145/3803437.3805575}
\acmISBN{979-8-4007-2636-1/2026/07}

\begin{document}

\title{PSR\textsuperscript{2}: A Phase-based Semantic Reasoning Framework for Atomicity Violation Detection via Contract Refinement}


\author{Xin Wang}
\email{24210839000020@hainanu.edu.cn}
\affiliation{%
  \institution{Hainan University}
  \city{Haikou}
  \country{China}
}

\author{Wenkai Li}
\email{cswkli@hainanu.edu.cn}
\affiliation{%
  \institution{Hainan University}
  \city{Haikou}
  \country{China}
}

\author{Zongwei Li}
\email{lizw1017@hainanu.edu.cn}
\affiliation{%
  \institution{Hainan University}
  \city{Haikou}
  \country{China}
}

\author{Xiaoqi Li}
\email{csxqli@ieee.org}
\affiliation{%
  \institution{Hainan University}
  \city{Haikou}
  \country{China}
}


\begin{abstract}
With the rapid advancement of decentralized applications, smart contract security faces severe challenges, particularly regarding atomicity violations in complex logic such as Oracle and NFT contracts. Rigid rule sets often limit traditional static analyzers and lack deep contextual awareness, leading to high false-positive and false-negative rates when identifying vulnerabilities that depend on intermediate state inconsistencies. To address these limitations, this paper proposes PSR\textsuperscript{2}, a novel collaborative static analysis framework that integrates structural path searching with deterministic semantic reasoning. PSR\textsuperscript{2} utilizes a Graph Structure Analysis Module (GSAM) to identify suspicious execution sequences in control flow graphs and a Semantic Context Analysis Module (SCAM) to extract data dependencies and state facts from abstract syntax trees. A Fusion Decision Module (FDM) then performs formal cross validation to confirm vulnerabilities based on a unified atomicity inconsistency model. Experimental results on 1,600 contract samples demonstrate that PSR\textsuperscript{2} significantly outperforms pattern-matching baselines, achieving an F1-score of 92.36\% in complex ERC-721 scenarios compared to 51.86\% for existing tools. Ablation studies further confirm that our fusion logic effectively reduces the false-positive rate by nearly half compared to single module analysis.
\end{abstract}



\maketitle

\vspace{-3mm}
\section{Introduction}
Oracles~\cite{caldarelli2025can,eskandari2021sok} serve as critical bridges for external data integration within decentralized finance~\cite{xi2024pomabuster}. However, the complexity of cross-domain interactions makes them highly susceptible to atomicity violations~\cite{tao2023atomicity,badruddoja2021making}. Traditional security analysis relies heavily on static analysis tools like Slither~\cite{feist2019slither} and pattern matching engines like Semgrep~\cite{durieux2020empirical}. While these approaches can identify surface-level vulnerabilities, they struggle with the deep state consistency and operation sequencing required in complex Oracle and NFT contracts~\cite{varma2025nft}. This inadequacy fundamentally stems from two limitations:\par

\noindent\textbf{Challenge 1 (C1): \textit{Contextual Blindness}.} Tools based on rigid rule sets often lack deep semantic awareness of variable dependencies, leading to high false positive rates by failing to distinguish benign updates~\cite{khan2022empirical} from malicious exploitations~\cite{sun2024gptscan}.\par

\noindent\textbf{Challenge 2 (C2): \textit{Topological Insensitivity}.} Pattern matching approaches focus on local code syntax but often fail to verify if a hazardous pattern is reachable via a valid control flow path. This limitation challenges the ability to detect implicit vulnerabilities (e.g., in ERC-721 callbacks) where the risk is embedded in the global execution topology rather than local syntax~\cite{duy2025vulnsense}, resulting in false negatives.\par

To address these challenges, we propose PSR$^2$, a collaborative static analysis framework. PSR$^2$ integrates structural path searching with deterministic semantic reasoning~\cite{guo2024enhanced}. By fusing graph-based evidence with semantic facts, our framework effectively eliminates structural ambiguities, ultimately enhancing both the precision and recall of vulnerability detection.\par

\noindent\textbf{Solution to C1:} PSR\textsuperscript{2} employs a Semantic Context Analysis Module (SCAM) to parse the smart contract into an Abstract Syntax Tree (AST)~\cite{tikhomirov2018smartcheck}. SCAM extracts deterministic semantic facts, such as data dependency sets and function roles, providing a rigorous logical foundation to filter out contextually invalid risks.\par

\noindent\textbf{Solution to C2:} PSR\textsuperscript{2} framework utilizes a Graph Structure Analysis Module (GSAM) to construct a simplified Control Flow Graph (CFG) and identify reachable hazardous paths (e.g., Read-Call-Write sequences)~\cite{qian2020towards}. A Fusion Decision Module (FDM) then cross validates these paths against semantic facts to resolve topological ambiguity. This prunes structurally valid but semantically infeasible paths. \par

The main contributions of this paper are as follows:
\begin{itemize}
    \item We propose PSR\textsuperscript{2}, a novel framework that harmonizes path-sensitive graph analysis with context-aware semantic extraction. It effectively resolves the trade-off between false positives and false negatives inherent in traditional tools.
    \item We define a Unified Atomicity Inconsistency Model that abstracts diverse vulnerabilities into a consistent logic primitive. Extensive experiments on 1,600 samples demonstrate that PSR\textsuperscript{2} outperforms traditional static analysis tools, achieving a 92.36\% F1-score in complex ERC-721 scenarios~\cite{das2022understanding}.
    \item We design a Fusion Decision Module (FDM) that aligns structural paths with semantic data flows for accurate and interpretable vulnerability detection.
\end{itemize}
\begin{figure*}[t]
    \centering
    \includegraphics[width=1\linewidth]{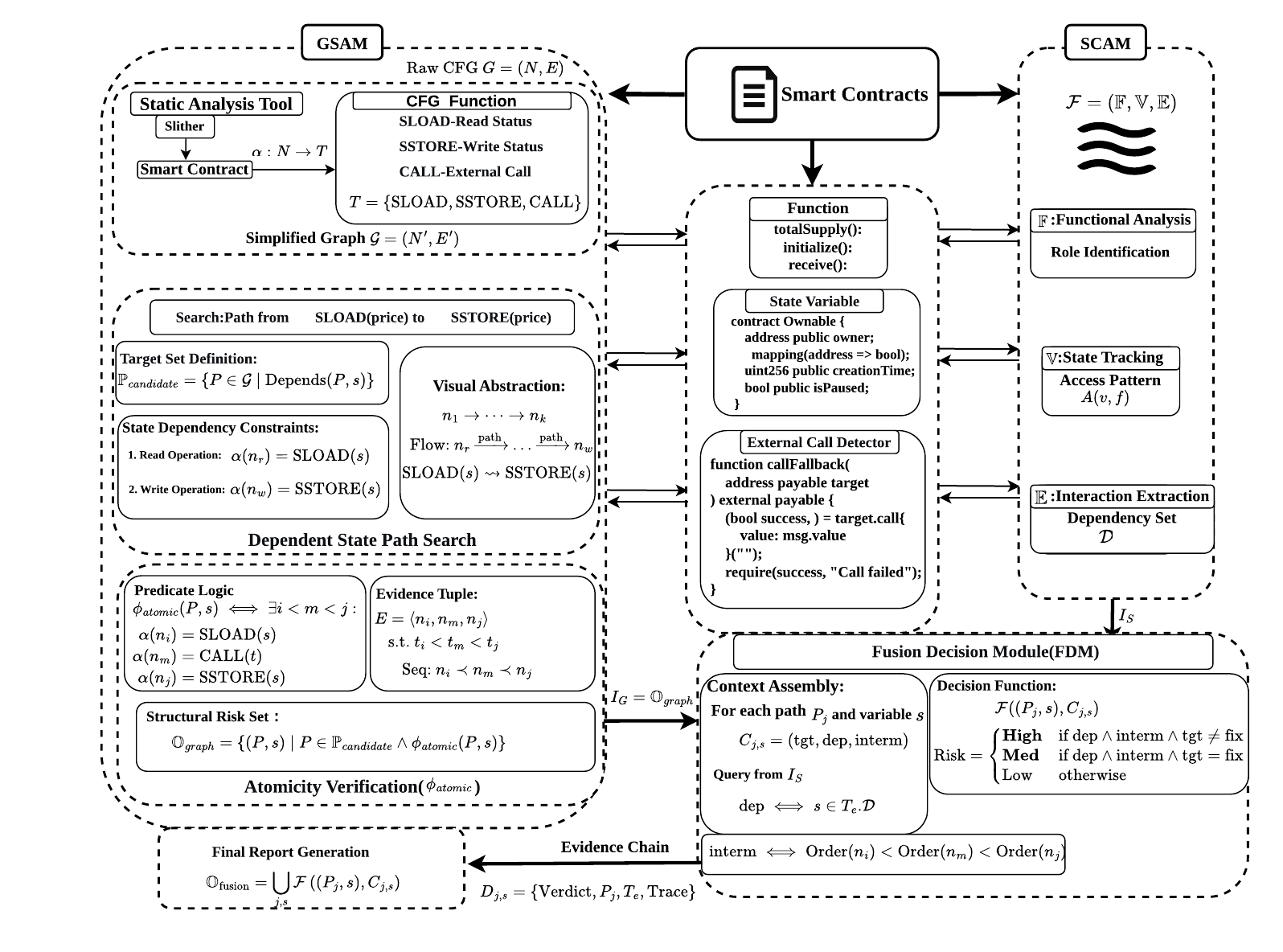}
    \caption{Architecture of the PSR$^2$ framework. It integrates parallel analysis from GSAM and SCAM modules, followed by a collaborative fusion decision for final vulnerability assessment.}
    \label{framework}
\end{figure*}
\section{Method}
As shown in Fig. \ref{framework}, $PSR^2$ involves three stages: (1) Semantic Context Analysis Module (SCAM): Parses source code $C$ into semantic facts. (2) Graph Structure Analysis Module (GSAM): Searches the Control Flow Graph (CFG) for hazardous paths (e.g., ``SLOAD$\to$CALL$\to\\$SSTORE''). (3) Fusion Decision Module (FDM): Cross-validates GSAM and SCAM findings to filter false positives and confirm atomicity violations~\cite{li2024static,luo2023two}.
\vspace{-3mm}
\subsection{SCAM}
This stage establishes the logical ground truth by parsing the contract $C$ into a deterministic semantic repository $\mathcal{F}$, eliminating probabilistic guessing:
\vspace{-2mm}
\[
\mathcal{F} = (\mathbb{F}, \mathbb{V}, \mathbb{E})
\]
\noindent \textbf{Function and State Profiling ($\mathbb{F}, \mathbb{V}$):}
SCAM first maps the contract's static topology. For every function $f$ and state variable $v$, it derives formal descriptors to categorize logical roles and access patterns:
\[
\begin{aligned}
D_f &= (\mathrm{id}, \mathrm{name}, \mathrm{role}) \\
D_v &= (\mathrm{id}, \mathrm{name}, \mathrm{type}, \{(f, A(v, f))\})
\end{aligned}
\]
Here, $\mathrm{role}$ is assigned by matching signatures against a keyword set $\Sigma = \{\text{price, update, \dots}\}$, and $A(v, f) \in \{\mathsf{read, write, RW}\}$ is determined via AST traversal. The resulting set $\mathbb{V} = \{D_v\}$ serves as the baseline for global dependency tracking.

\noindent \textbf{Interaction Dependency Extraction ($\mathbb{E}$):}
To capture atomicity risks, SCAM analyzes every external call site $e$, generating a critical fact tuple $T_e$:
\[
T_e = (\text{loc, type, tgt, } \mathcal{D})
\]
Where $\text{loc}$ denotes code coordinates, and $\text{tgt}$ classifies the target source (e.g., fixed address vs. user input). Crucially, SCAM performs data flow analysis to construct the \textbf{dependency set} $\mathcal{D} \subseteq \mathbb{V}$ ~\cite{ghaleb2022towards,wang2024smart}, which identifies specifically which state variables dictate the call's parameters or execution conditions.
\vspace{-3mm}
\subsection{GSAM}
The GSAM transforms the contract's CFG into a formal model to verify path reachability and operation sequencing. It outputs a set of suspicious paths $\mathbb{P}_{\text{suspicious}}$ satisfying risk patterns, providing structural evidence for the FDM.

\noindent \textbf{Graph Construction and Abstraction:}
Instead of analyzing raw basic blocks, GSAM projects the contract into a simplified graph $\mathcal{G} = (N', E')$ using a labeling function $\alpha: N \to T$~\cite{wang2025contractscanner,cao2023data}. This function maps nodes to critical atomicity related operations, filtering out irrelevant logic:
\[
T = \left\{
\begin{aligned}
&\mathrm{SLOAD}(s), \mathrm{SSTORE}(s), \\
&\mathrm{CALL}(t), \mathrm{CHECK}, \mathrm{OTHER}
\end{aligned}
\ \middle| \ 
\begin{aligned}
&s \in \mathbb{V}, \\
&t \in \mathrm{AddressType}
\end{aligned}
\right\}
\]
The simplified graph nodes are defined as $N' = \{\alpha(n) \mid n \in N\}$, preserving the original control flow edges $E'$.

\noindent \textbf{Atomicity Pattern Verification:}
GSAM identifies "Dependent State Paths" that contain both read and write operations on a critical variable $s$. For any such path $P = \langle n_1, \dots, n_k \rangle$, the module evaluates the atomicity safety predicate $\phi_{\text{atomic}}(P, s)$. A violation is flagged if an external call interrupts the read write sequence~\cite{zhang2023demystifying,wu2024advscanner}:
\[
\begin{aligned}
\phi_{\text{atomic}}(P, s) \iff \exists i, m, j \in [1, k] : \quad & (i < m < j) \\
& \land (\alpha(n_i) = \mathrm{SLOAD}(s)) \\
& \land (\alpha(n_m) = \mathrm{CALL}(t)) \\
& \land (\alpha(n_j) = \mathrm{SSTORE}(s))
\end{aligned}
\]
The final output is a repository of structurally risky paths, defined as:
\[
\mathbb{O}_{\text{graph}} = \left\{ (\mathcal{R}, P, s, \text{Seq}) \mid P \in \mathbb{P}_{\text{candidate}} \land \phi_{\text{atomic}}(P, s) \right\}
\]
where $\text{Seq}$ denotes the evidence node sequence.
\vspace{-2mm}
\subsection{FDM}
Operating as the logical gatekeeper, the FDM synthesizes structural alerts $I_G$ from GSAM and semantic facts $I_S$ from SCAM. 

\noindent \textbf{Contextual Analysis:}
For each suspicious path report $R_j \in I_G$ involving state variable $s$, the FDM constructs a semantic context annotation $C_{j,s} = (\text{tgt}, \text{dep}, \text{interm})$ by querying $I_S$:
\begin{itemize}
    \item $\text{tgt}$: The call target type retrieved from facts $T_e$.
    \item $\text{dep}$: Boolean flag, true iff $s \in T_e.\mathcal{D}$ (verifying data dependency).
    \item $\text{interm}$: Boolean flag, true iff the call occurs strictly between $\mathrm{SLOAD}(s)$ and $\mathrm{SSTORE}(s)$ (verifying intermediate state inconsistency).
\end{itemize}

\noindent \textbf{Fused Decision Logic:}
The module executes a deterministic decision function $\mathcal{F}$ to assign risk levels. A vulnerability is confirmed only when structural reachability is cross-validated by semantic dependencies:
\[
\mathcal{F}\big((P_j,s), C_{j,s}\big) =
\begin{cases}
(\text{High}, D_{j,s}) & \text{if } \text{dep} \land \text{interm} \land \text{tgt} \neq \text{fixed} \\
(\text{Medium}, D_{j,s}) & \text{if } \text{dep} \land \text{interm} \land \text{tgt} = \text{fixed} \\
(\text{Low}, \emptyset) & \text{otherwise}
\end{cases}
\]
If the risk is High or Medium, a structured evidence report $D_{j,s}$ is synthesized, encapsulating the verification chain $\{ \text{Verdict}, P_j, s, T_e, \\C_{j,s} \}$.

\noindent \textbf{Final Output:}
The system yields a comprehensive security report $\mathbb{O}_{\text{fusion}}$, representing the union of all validated atomicity violations:
\vspace{-2mm}
\[
\mathbb{O}_{\text{fusion}} = \bigcup_{j, s} \mathcal{F}\left( (P_{j}, s), C_{j,s} \right)
\]
\vspace{-2mm}
\vspace{-5mm}
\section{Experiment}
\label{sec:experiment}

\noindent\textbf{Experimental Setup \& Baselines.}
We evaluate PSR\textsuperscript{2} on a comprehensive dataset of 1,600 samples across three benchmarks: {Smartbugs-Reentrancy}, {ERC-721 Reentrancy} (featuring complex implicit callbacks), and {Smartbugs-Unchecked External Call}.\par
We compare PSR\textsuperscript{2} against two state-of-the-art industry tools: {Slither} (static analysis) and {Semgrep} (pattern matching). Our evaluation addresses three key questions: \underline{RQ1:} How does PSR\textsuperscript{2} compare to baselines in detection accuracy? \underline{RQ2:} What is the specific contribution of each module to the synergy? \underline{RQ3:} How generalizable is the unified model across diverse risks?

\begin{table}[h]
\centering
\small 
\setlength{\tabcolsep}{3.9pt} 
\caption{Performance Comparison across Multiple Datasets%
\textsuperscript{\ref{fn:corrigendum}}}
\vspace{-2mm}
\label{tab:performance_comparison}
\begin{tabular}{llccc}
\hline
\textbf{Dataset} & \textbf{Tool} & \textbf{Precision (\%)} & \textbf{Recall (\%)} & \textbf{F1-score (\%)} \\ \hline
\multirow{3}{*}{Reentrancy} & Slither & 12.50 & 2.94 & 4.76 \\
 & Semgrep & 71.32 & 95.10 & 81.51 \\
 & \textbf{PSR\textsuperscript{2} } & \textbf{78.13} & \textbf{98.04} & \textbf{86.96} \\ \hline
\multirow{3}{*}{ERC-721 Reent.} & Slither & 16.31 & 21.45 & 18.53 \\
 & Semgrep & 59.90 & 45.73 & 51.86 \\
 & \textbf{PSR\textsuperscript{2} } & \textbf{94.94} & \textbf{89.91} & \textbf{92.36} \\ \hline
\multirow{3}{*}{Unchecked} & Slither & 74.12 & 71.93 & 73.01 \\
 & Semgrep & 55.00 & 9.65 & 16.42 \\
 & \textbf{PSR\textsuperscript{2} } & \textbf{84.21} & \textbf{98.25} & \textbf{90.69} \\ \hline
\end{tabular}
\vspace{-1.5em}
\end{table}
\vspace{-3mm}
\subsection{Performance Analysis (RQ1)}
As shown in Table \ref{tab:performance_comparison}, PSR\textsuperscript{2} consistently achieves the highest F1-scores across all datasets, demonstrating significant advantages in both Precision and Recall.

\textbf{Handling Complex Callbacks (ERC-721):} PSR\textsuperscript{2} achieves a dominant F1-score of 92.36\%, far surpassing Semgrep (51.86\%) and Slither (18.53\%). Slither suffers from "Contextual Blindness" (C1), failing to trace control flows when external calls trigger implicit callbacks (e.g., \texttt{onERC721Received})~\cite{xue2020cross}. Conversely, Semgrep is limited by topological insensitivity, merely matching local syntax without verifying strict reachability. PSR\textsuperscript{2} overcomes these limitations by effectively fusing topological searching (GSAM) with semantic verification (SCAM).

\textbf{Generalizability:} On the Unchecked dataset, PSR\textsuperscript{2} maintains a high recall of 98.25\%, whereas Semgrep drops drastically to 9.65\%. This sharp decline reveals the fragility of pattern based detection relying on rigid rules~\cite{jeon2024design}. Driven by a unified "Atomic Inconsistency" model rather than fixed signatures, PSR\textsuperscript{2} demonstrates superior robustness across diverse and non-standard implementation styles.
\vspace{-5mm}
\begin{tcolorbox}[boxrule=1pt,boxsep=4pt,left=1pt,right=1pt,top=0.5pt,bottom=0.5pt, colback=gray!5, after skip=-3pt]
\textbf{Answer to RQ1:} PSR\textsuperscript{2} significantly outperforms existing baselines. By effectively coupling structural path searching with deep semantic reasoning, it eliminates the context blindness and pattern rigidity inherent in traditional static analyzers, ensuring high accuracy and generalizability.
\end{tcolorbox}
\subsection{Ablation Study (RQ2)}

To evaluate the synergy of multi-module fusion, we compare the full framework against two decoupled variants: \textbf{$PSR^2_{GSAM}$} (Graph-only, strictly structural analysis without semantic confirmation) and \textbf{$PSR^2_{SCAM}$} (Semantic-only, pattern matching without path verification). To mitigate class imbalance bias, we evaluated on a strictly balanced dataset comprising 719 vulnerable ($D_{vuln}$) and 719 safe ($D_{safe}$) contracts.
\vspace{-2mm}
\begin{table}[h]
\vspace{-2mm}
\centering
\small 
\setlength{\tabcolsep}{5pt} 
\caption{Ablation Metrics on Decoupled Datasets%
\tablefootnote[1]{%
We identified a few typographical and layout errors in Tables 1 and 2 due to manual math conversion error and LaTeX table column misalignment. The paper's core findings and conclusions are completely unaffected. We have submitted a formal Corrigendum to update errors in the ACM Digital Library.
\label{fn:corrigendum}
}}
\vspace{-3mm}
\label{tab:ablation_study}
\begin{tabular}{lccccc}
\hline
\multirow{2}{*}{\textbf{Configuration}} & \multicolumn{2}{c}{\textbf{$D_{vuln}$ (Rec.)}} & \multicolumn{2}{c}{\textbf{$D_{safe}$ (FPR)}} & \multirow{2}{*}{\textbf{Over. $F_1$}} \\ \cline{2-5}
 & Count & \% & Count & \% & \\ \hline
$PSR^2_{GSAM}$ & 639 & 88.87\% & 81 & 11.27\% & 88.81\% \\
$PSR^2_{SCAM}$ & 602 & 83.73\% & 594 & 82.61\% & 62.87\% \\
\textbf{$PSR^2_{Full}$} & \textbf{664} & \textbf{92.35\%} & \textbf{46} & \textbf{6.40\%} & \textbf{92.93\%} \\ \hline
\end{tabular}
\vspace{-2mm}
\end{table}

As shown in Table \ref{tab:ablation_study}, decoupled modules suffer from logical blindness. $PSR^2_{SCAM}$ exhibits extreme over reporting ($FPR=82.61\%$) as it flags patterns regardless of execution feasibility. Similarly, $PSR^2_{GSAM}$ generates structural noise ($FPR=11.27\%$) on benign paths.
\par
The FDM effectively prunes noise by requiring semantic "Facts" (e.g., state inconsistency) to cross-validate structural paths, reducing the FPR to \textbf{6.40\%}. Furthermore, the fusion logic resolves the "implicit callback" blindness, elevating the recall to 92.35\% by identifying complex ERC-721 vulnerabilities that rely on both structural reachability and semantic triggers.

\begin{tcolorbox}[boxrule=1pt,boxsep=4pt,left=1pt,right=1pt,top=0.5pt,bottom=0.5pt, colback=gray!5, after skip=-3pt]
\textbf{Answer to RQ2:} Ablation confirms the necessity of fusing path sensitivity (GSAM) with semantic context (SCAM). FDM filters structural noise and captures implicit vulnerabilities that decoupled modules miss.
\end{tcolorbox}

\subsection{Generality of the Unified Model (RQ3)}
{CrossCategory Primitive Extraction:}
The core innovation is refining raw code into a unified "Atomic Inconsistency" primitive. Our evaluation reveals that diverse vulnerabilities share the same root logic:

{Reentrancy} (explicit and implicit) is refined as a structural violation sequence $\langle \mathrm{SLOAD}(s), \mathrm{CALL}(t), \mathrm{SSTORE}(s) \rangle$, confirming the breach of the "Check-Effects-Interactions" pattern.

{Unchecked External Calls} are treated as logic-state inconsistencies where execution proceeds to $\mathrm{SSTORE}$ without validation from the call's return value ($T_e$).
Crucially, both categories share the same pathological root: a critical variable resides in an "Intermediate State" (loaded but not yet committed) during a side-effect-heavy interaction~\cite{li2025understanding}.

\noindent{Unified Decision Logic.}
The FDM applies a standardized decision function $\mathcal{F}$ regardless of the specific vulnerability signature. It confirms risks only when structural reachability is cross-validated by semantic dependencies ($E_S.\text{dep}$).This logical consistency facilitates the high performance observed in Table~\ref{tab:performance_comparison}. This generality is most evident in ERC-721 scenarios, where PSR\textsuperscript{2} uses semantic reasoning to complete broken control flows (implicit paths), identifying violations that are structurally hidden to traditional tools but semantically certain.

\begin{tcolorbox}[boxrule=1pt,boxsep=4pt,left=1pt,right=1pt,top=0.5pt,bottom=0.5pt, colback=gray!5 ]
\textbf{Answer to RQ3:}
PSR\textsuperscript{2} achieves high generalizability by abstracting diverse bugs into a unified triplet $(P_j, T_e, D_s)$. Through semantic reasoning, the FDM consistently detects "Atomicity Violations", effectively transcending traditional rule-based limitations.
\end{tcolorbox}
\vspace{-3mm}
\section{DISCUSSION}
PSR\textsuperscript{2} achieves high precision by utilizing the FDM as a logical gatekeeper~\cite{wei2025advanced} to filter structural noise through the cross-validation of GSAM’s paths and SCAM’s semantic facts. This synergy allows the detection of ``structurally hidden'' reentrancy in ERC-721 callbacks via interaction dependency extraction, addressing the limitations of pattern matching~\cite{song2025silence}. However, mapping complete dependency sets $\mathcal{D}$ remains challenging for cross-contract calls without source code~\cite{mishra2025blockchain}, a domain where semantic lifting at the bytecode level becomes essential~\cite{grech2022elipmoc}. By refining diverse risks into structured primitives $(P_j, T_e, D_s)$, the unified model transcends manual rule-writing and establishes a scalable foundation for securing evolving DeFi protocols against sophisticated cross-contract exploits.

\vspace{-2mm}
\section{Related Work}
Static analysis utilizes intermediate representations to verify program safety~\cite{tsankov2018securify}, but faces path explosion and high false-positive rates due to limited semantic context. Conversely, pattern matching provides computational efficiency~\cite{zheng2024dappscan} but lacks topological sensitivity for implicit control flows like cross-function callbacks. Despite recent graph-based advances in transaction network monitoring, existing frameworks still struggle to unify structural reachability with semantic dependency~\cite{liu2025detecting} often missing atomicity violations that require simultaneous path and state verification~\cite{trozze2024detecting}.
\section{CONCLUSION}
We propose PSR\textsuperscript{2}, a framework integrating path-sensitive graph analysis with context-aware semantic extraction. Its multi-module architecture (SCAM, GSAM, and FDM) effectively resolves the precision-recall tradeoff. Experiments show a 92.36\% F1-score in complex ERC-721 scenarios, significantly outperforming state-of-the-art tools~\cite{xu2025enhanced}. This work establishes a unified approach to securing decentralized systems against atomicity violations~\cite{gogol2024sok}, ensuring deep state consistency and supporting broader blockchain applications, such as IoT data sharing.
\vspace{-2mm}
\section{ACKNOWLEDGMENTS}
We acknowledge the use of AI for grammar improvement and text polishing.

\appendix
\balance
\bibliographystyle{ACM-Reference-Format}
\bibliography{references}

@article{caldarelli2025can,
  title={Can artificial intelligence solve the blockchain oracle problem? unpacking the challenges and possibilities},
  author={Caldarelli, Giulio},
  journal={Frontiers in Blockchain},
  volume={8},
  pages={1682623},
  year={2025},
  publisher={Frontiers Media SA}
}

@inproceedings{feist2019slither,
  title={Slither: a static analysis framework for smart contracts},
  author={Feist, Josselin and Grieco, Gustavo and Groce, Alex},
  booktitle={2019 IEEE/ACM 2nd International Workshop on Emerging Trends in Software Engineering for Blockchain (WETSEB)},
  pages={8--15},
  year={2019},
  organization={IEEE}
}

@inproceedings{xi2024pomabuster,
  title={POMABuster: Detecting Price Oracle Manipulation Attacks in Decentralized Finance},
  author={Xi, Rui and Wang, Zehua and Pattabiraman, Karthik},
  booktitle={2024 IEEE Symposium on Security and Privacy (SP)},
  pages={3923--3942},
  year={2024},
  organization={IEEE}
}

@article{tao2023atomicity,
  title={On atomicity and confidentiality across blockchains under failures},
  author={Tao, Yuechen and Li, Bo and Li, Baochun},
  journal={IEEE Transactions on Knowledge and Data Engineering},
  volume={36},
  number={2},
  pages={766--780},
  year={2023},
  publisher={IEEE}
}

@inproceedings{durieux2020empirical,
  title={Empirical review of automated analysis tools on 47,587 ethereum smart contracts},
  author={Durieux, Thomas and Ferreira, Jo{\~a}o F and Abreu, Rui and Cruz, Pedro},
  booktitle={Proceedings of the ACM/IEEE 42nd International conference on software engineering},
  pages={530--541},
  year={2020}
}

@article{varma2025nft,
  title={NFT Marketplaces: A Comprehensive Analysis of Trading, Security, and Metadata Challenges},
  author={Varma, Shrey and Prajapati, Sachin and Gupta, YashKumar and Tondon, Kaushik and Sharma, Shraddha and Parate, Manali and Churi, Manasi},
  journal={International Journal on Advanced Electrical and Computer Engineering},
  volume={14},
  number={1},
  pages={55--68},
  year={2025}
}

@inproceedings{sun2024gptscan,
  title={Gptscan: Detecting logic vulnerabilities in smart contracts by combining gpt with program analysis},
  author={Sun, Yuqiang and Wu, Daoyuan and Xue, Yue and Liu, Han and Wang, Haijun and Xu, Zhengzi and Xie, Xiaofei and Liu, Yang},
  booktitle={Proceedings of the IEEE/ACM 46th International Conference on Software Engineering},
  pages={1--13},
  year={2024}
}

@article{duy2025vulnsense,
  title={Vulnsense: Efficient vulnerability detection in ethereum smart contracts by multimodal learning with graph neural network and language model},
  author={Duy, Phan The and Khoa, Nghi Hoang and Quyen, Nguyen Huu and Trinh, Le Cong and Kien, Vu Trung and Hoang, Trinh Minh and Pham, Van-Hau},
  journal={International Journal of Information Security},
  volume={24},
  number={1},
  pages={48},
  year={2025},
  publisher={Springer}
}

@article{wei2025advanced,
  title={Advanced smart contract vulnerability detection via llm-powered multi-agent systems},
  author={Wei, Zhiyuan and Sun, Jing and Sun, Yuqiang and Liu, Ye and Wu, Daoyuan and Zhang, Zijian and Zhang, Xianhao and Li, Meng and Liu, Yang and Li, Chunmiao and others},
  journal={IEEE Transactions on Software Engineering},
  year={2025},
  publisher={IEEE}
}

@article{guo2024enhanced,
  title={An enhanced state-aware model learning approach for security analysis in lightweight protocol implementations},
  author={Guo, Jiaxing and Zhao, Dongliang and Gu, Chunxiang and Chen, Xi and Zhang, Xieli and Ju, Mengcheng},
  journal={Journal of Cloud Computing},
  volume={13},
  number={1},
  pages={28},
  year={2024},
  publisher={Springer}
}

@inproceedings{tsankov2018securify,
  title={Securify: Practical security analysis of smart contracts},
  author={Tsankov, Petar and Dan, Andrei and Drachsler-Cohen, Dana and Gervais, Arthur and Buenzli, Florian and Vechev, Martin},
  booktitle={Proceedings of the 2018 ACM SIGSAC conference on computer and communications security},
  pages={67--82},
  year={2018}
}

@article{zheng2024dappscan,
  title={Dappscan: building large-scale datasets for smart contract weaknesses in dapp projects},
  author={Zheng, Zibin and Su, Jianzhong and Chen, Jiachi and Lo, David and Zhong, Zhijie and Ye, Mingxi},
  journal={IEEE Transactions on Software Engineering},
  volume={50},
  number={6},
  pages={1360--1373},
  year={2024},
  publisher={IEEE}
}

@article{wang2025contractscanner,
  title={ContractScanner: Detecting and Localizing Vulnerabilities of Smart Contracts via Graph-Based Semantic Modeling of Source Code},
  author={Wang, Bin and Li, Shan and Yuan, Xiaohan and Xie, Xueshuo and Wang, Junyong and Li, Tao and Wang, Wei},
  journal={IEEE Transactions on Network Science and Engineering},
  year={2025},
  publisher={IEEE}
}

@article{trozze2024detecting,
  title={Detecting DeFi securities violations from token smart contract code},
  author={Trozze, Arianna and Kleinberg, Bennett and Davies, Toby},
  journal={Financial Innovation},
  volume={10},
  number={1},
  pages={1--35},
  year={2024},
  publisher={Springer}
}

@inproceedings{song2025silence,
  title={Silence False Alarms: Identifying Anti-Reentrancy Patterns on Ethereum to Refine Smart Contract Reentrancy Detection},
  author={Song, Qiyang and Huang, Heqing and Jia, Xiaoqi and Xie, Yuanbo and Cao, Jiahao},
  booktitle={NDSS},
  year={2025}
}

@article{mishra2025blockchain,
  title={Blockchain Security in Focus: A Comprehensive Investigation into Threats, Smart Contract Security, Cross-Chain Bridges, Vulnerabilities Detection Tools \& Techniques},
  author={Mishra, Deepa and Phansalkar, Shraddha},
  journal={IEEE Access},
  year={2025},
  publisher={IEEE}
}

@article{xu2025enhanced,
  title={Enhanced Smart Contract Vulnerability Detection via Graph Neural Networks: Achieving High Accuracy and Efficiency},
  author={Xu, Chang and Xu, Huaiyu and Zhu, Liehuang and Shen, Xiaodong and Sharif, Kashif},
  journal={IEEE Transactions on Software Engineering},
  year={2025},
  publisher={IEEE}
}

@inproceedings{ghaleb2022towards,
  title={Towards effective static analysis approaches for security vulnerabilities in smart contracts},
  author={Ghaleb, Asem},
  booktitle={Proceedings of the 37th IEEE/ACM International Conference on Automated Software Engineering},
  pages={1--5},
  year={2022}
}

@inproceedings{zhang2023demystifying,
  title={Demystifying exploitable bugs in smart contracts},
  author={Zhang, Zhuo and Zhang, Brian and Xu, Wen and Lin, Zhiqiang},
  booktitle={2023 IEEE/ACM 45th International Conference on Software Engineering (ICSE)},
  pages={615--627},
  year={2023},
  organization={IEEE}
}

@article{qian2020towards,
  title={Towards automated reentrancy detection for smart contracts based on sequential models},
  author={Qian, Peng and Liu, Zhenguang and He, Qinming and Zimmermann, Roger and Wang, Xun},
  journal={IEEE access},
  volume={8},
  pages={19685--19695},
  year={2020},
  publisher={IEEE}
}

@inproceedings{badruddoja2021making,
  title={Making smart contracts smarter},
  author={Badruddoja, Syed and Dantu, Ram and He, Yanyan and Upadhayay, Kritagya and Thompson, Mark},
  booktitle={2021 IEEE International Conference on Blockchain and Cryptocurrency (ICBC)},
  pages={1--3},
  year={2021},
  organization={IEEE}
}

@inproceedings{eskandari2021sok,
  title={Sok: Oracles from the ground truth to market manipulation},
  author={Eskandari, Shayan and Salehi, Mehdi and Gu, Wanyun Catherine and Clark, Jeremy},
  booktitle={Proceedings of the 3rd ACM Conference on Advances in Financial Technologies},
  pages={127--141},
  year={2021}
}

@article{khan2022empirical,
  title={Empirical analysis of vulnerabilities in blockchain-based smart contracts},
  author={Khan, Kashif Mehboob and Zahid, Ansha},
  journal={Sir Syed University Research Journal of Engineering \& Technology},
  volume={12},
  number={1},
  pages={78--85},
  year={2022}
}

@inproceedings{tikhomirov2018smartcheck,
  title={Smartcheck: Static analysis of ethereum smart contracts},
  author={Tikhomirov, Sergei and Voskresenskaya, Ekaterina and Ivanitskiy, Ivan and Takhaviev, Ramil and Marchenko, Evgeny and Alexandrov, Yaroslav},
  booktitle={Proceedings of the 1st international workshop on emerging trends in software engineering for blockchain},
  pages={9--16},
  year={2018}
}

@inproceedings{das2022understanding,
  title={Understanding security issues in the NFT ecosystem},
  author={Das, Dipanjan and Bose, Priyanka and Ruaro, Nicola and Kruegel, Christopher and Vigna, Giovanni},
  booktitle={Proceedings of the 2022 ACM SIGSAC conference on computer and communications security},
  pages={667--681},
  year={2022}
}

@article{li2024static,
  title={Static application security testing (sast) tools for smart contracts: How far are we?},
  author={Li, Kaixuan and Xue, Yue and Chen, Sen and Liu, Han and Sun, Kairan and Hu, Ming and Wang, Haijun and Liu, Yang and Chen, Yixiang},
  journal={Proceedings of the ACM on Software Engineering},
  volume={1},
  number={FSE},
  pages={1447--1470},
  year={2024},
  publisher={ACM New York, NY, USA}
}

@inproceedings{luo2023two,
  title={Two-Stage Smart Contract Vulnerability Detection Combining Semantic Features and Graph Features},
  author={Luo, Zhenkun and Chen, Shuhong and Wang, Guojun and Li, Hanjun},
  booktitle={2023 IEEE 22nd International Conference on Trust, Security and Privacy in Computing and Communications (TrustCom)},
  pages={257--264},
  year={2023},
  organization={IEEE}
}

@inproceedings{wang2024smart,
  title={Smart Contract Vulnerability Detection via Feature Fusion of Local Data Flow and Global Features},
  author={Wang, Long and Chen, Zhihua and Pang, Hua and Li, Xiaoguang},
  booktitle={2024 IEEE International Symposium on Parallel and Distributed Processing with Applications (ISPA)},
  pages={2268--2271},
  year={2024},
  organization={IEEE}
}

@inproceedings{cao2023data,
  title={Data flow-driven and attention mechanism-enabled smart contract vulnerability detection for secure and green blockchain-based service networks},
  author={Cao, Yuanlong and Jiang, Fan and Xiao, Jianmao and Chen, Shaolong and Yang, Wei and Yi, Yugen},
  booktitle={ICC 2023-IEEE International Conference on Communications},
  pages={5135--5140},
  year={2023},
  organization={IEEE}
}

@inproceedings{wu2024advscanner,
  title={Advscanner: Generating adversarial smart contracts to exploit reentrancy vulnerabilities using llm and static analysis},
  author={Wu, Yin and Xie, Xiaofei and Peng, Chenyang and Liu, Dijun and Wu, Hao and Fan, Ming and Liu, Ting and Wang, Haijun},
  booktitle={Proceedings of the 39th IEEE/ACM International Conference on Automated Software Engineering},
  pages={1019--1031},
  year={2024}
}

@inproceedings{xue2020cross,
  title={Cross-contract static analysis for detecting practical reentrancy vulnerabilities in smart contracts},
  author={Xue, Yinxing and Ma, Mingliang and Lin, Yun and Sui, Yulei and Ye, Jiaming and Peng, Tianyong},
  booktitle={Proceedings of the 35th IEEE/ACM International Conference on Automated Software Engineering},
  pages={1029--1040},
  year={2020}
}

@article{jeon2024design,
  title={Design and evaluation of highly accurate smart contract code vulnerability detection framework},
  author={Jeon, Sowon and Lee, Gilhee and Kim, Hyoungshick and Woo, Simon S},
  journal={Data Mining and Knowledge Discovery},
  volume={38},
  number={3},
  pages={888--912},
  year={2024},
  publisher={Springer}
}

@article{li2025understanding,
  title={Understanding inconsistent state update vulnerabilities in smart contracts},
  author={Li, Lantian and Chen, Yuyu and Wu, Jingwen and Pan, Yue and Yu, Zhongxing},
  journal={ACM Transactions on Software Engineering and Methodology},
  year={2025},
  publisher={ACM New York, NY}
}

@article{gogol2024sok,
  title={SoK: Decentralized Finance (DeFi)--Fundamentals, Taxonomy and Risks},
  author={Gogol, Krzysztof and Killer, Christian and Schlosser, Malte and Bocek, Thomas and Stiller, Burkhard and Tessone, Claudio},
  journal={arXiv preprint arXiv:2404.11281},
  year={2024}
}

@article{liu2025detecting,
  title={Detecting smart contract state-inconsistency bugs via flow divergence and multiplex symbolic execution},
  author={Liu, Yinxi and Meng, Wei and Zhang, Yinqian},
  journal={Proceedings of the ACM on Software Engineering},
  volume={2},
  number={FSE},
  pages={22--43},
  year={2025},
  publisher={ACM New York, NY, USA}
}

@article{grech2022elipmoc,
  title={Elipmoc: Advanced decompilation of ethereum smart contracts},
  author={Grech, Neville and Lagouvardos, Sifis and Tsatiris, Ilias and Smaragdakis, Yannis},
  journal={Proceedings of the ACM on Programming Languages},
  volume={6},
  number={OOPSLA1},
  pages={1--27},
  year={2022},
  publisher={ACM New York, NY, USA}
}

\end{document}